\magnification=1200

\def\a{\alpha}\def\d{\delta}\def\e{\epsilon}

\def\k{\kappa}\def\l{\lambda}\def\m{\mu}\def\n{\nu}\def
\p{\pi}\def\r{\rho}\def\s{\sigma}\def\t{\tau}
\def\y{\eta}\def\x{\xi}

\def\D{\Delta}\def\F{\Phi}\def\L{\Lambda}
\def\O{\Omega}\def\P{\Pi}\def\X{\Xi}

\def\de{\partial}
\def\inf{\infty}\def\id{\equiv}\def\mo{{-1}}\def\ha{{1\over 2}}

\def\mn{{\mu\nu}}

\def\fe{field equations }
\def\coo{coordinates }

\def\cosc{cosmological constant }

\def\pb{Poisson brackets }

\def\mi{Minkowski }\def\ads{anti-de Sitter }
\def\poi{Poincar\'e }
\def\des{de Sitter }

\def\GR{general relativity }

\def\wrt{with respect to }\def\tls{transformation laws }
\def\eom{equations of motion }

\def\section#1{\bigskip\noindent{\bf#1}\smallskip}
\def\subsect#1{\bigskip\noindent{\it#1}\smallskip}

\font\small = cmr8

\def\PL#1{Phys.\ Lett.\ {\bf#1}}
\def\PRL#1{Phys.\ Rev.\ Lett.\ {\bf#1}}
\def\PR#1{Phys.\ Rev.\ {\bf#1}}\def\CQG#1{Class.\ Quantum Grav.\ {\bf#1}}

\def\JMP#1{J.\ Math.\ Phys.\ {\bf#1}}

 \def\IJMP#1{Int.\ J. Mod.\ Phys.\ {\bf #1}}
\def\MPL#1{Mod.\ Phys.\ Lett.\ {\bf #1}} 

\def\AoP#1{Ann.\ Phys.\ {\bf#1}}
\def\hep#1{{\tt hep-th/#1}}

\def\ref#1{\medskip\everypar={\hangindent 2\parindent}#1}
\def\beginref{\begingroup
\bigskip
\centerline{\bf References}
\nobreak\noindent}
\def\endref{\par\endgroup}

\def\eps{\e_{ijk}}\def\Con{\left(1-{\L\tX^2\over4}\right)}

\def\til{\tilde}\def\tX{\til X}\def\tP{\til P}\def\tx{\til x}
\def\bX{\bar X}\def\bP{\bar P}\def\hX{\hat X}\def\hP{\hat P}
\def\dX{\dot X}\def\dP{\dot P}\def\cX{{\cal X}}\def\cP{{\cal P}}
\def\tp{\til p}\def\tx{\til x}\def\bcX{\bar{\cal X}}\def\bcP{\bar{\cal P}}
\def\cbX{\breve{\cal X}}\def\cbP{\breve{\cal P}}
\def\lox{1-\L\O\,\cbX^2\cbP^2}\def\op{1-\O\,\cbP^2}\def\lx{1-\L\,\cbX^2}
\def\opb{{1-\O\,\bar\cP^2}}
\def\hx{\hat x}\def\hp{\hat p}
\def\hfact{{(1-\hp_0/\k)^{-2}+\L\hx^2}}
\def\snyd{\sqrt{1+\O P^2}}\def\snyi{\sqrt{1-\O\cP^2}}
\def\trans{transformations }\def\pun{\cdotp\!}
\def\tls{transformation laws }
{\nopagenumbers
\line{\hfil January 2008}
\vskip80pt
\centerline{\bf Doubly special relativity in de Sitter spacetime}
\vskip40pt
\centerline{
{\bf S. Mignemi}\footnote{$^\ddagger$}{e-mail: smignemi@unica.it}}
\vskip10pt
\centerline {Dipartimento di Matematica, Universit\`a di Cagliari}
\centerline{viale Merello 92, 09123 Cagliari, Italy}
\centerline{and INFN, Sezione di Cagliari}
\vskip100pt
\centerline{\bf Abstract}

\vskip10pt
{\noindent
We discuss the generalization of Doubly Special Relativity to a curved \des
background. The model has three observer-independent scales, the velocity of
light $c$, the radius of curvature of the geometry $\a$, and the Planck energy
$\k$, and can be realized in a noncommutative position space.
It is possible to construct a model exhibiting a duality for the interchange
of positions and momenta together with the exchange of $\a$ and $\k$.
}
\vskip100pt\
P.A.C.S. Numbers: 11.30.Cp, 03.30.+p
\vfil\eject}

\section{1. Introduction}
Since the early years of general relativity,\des and \ads spaces have acquired a
fundamental importance, both theoretical and phenomenological, especially in the
context of cosmology.
Indeed, recent astrophysical observations seem to indicate that our universe
has positive \cosc [1].

In spite of their relevance, there is not much literature about the extension
of the kinematics of special relativity to \des or \ads
backgrounds\footnote{$^\dagger$}{\small To our knowledge, this is discussed only in
[2,3] for a specific choice of coordinates.}.
Geometrically, \des space is defined as a space of constant positive curvature.
Its isometries are generated by the \des algebra, that can be considered as a
deformation of the \poi algebra with a parameter $\a=1/\sqrt\L$ of
dimension of length.
Of course, several geometric and algebraic properties of \des space differ from
those of Minkowski space.
For example, contrary to flat space, in \des space the generators of the
translations cannot be identified with the canonical momenta, as is obvious
from the position dependence of the \des hamiltonian. Moreover, there is no
natural parametrization of the space and, depending on the specific problem
one is studying, different systems of coordinates can be more convenient.

A different kind of deformation of special relativity is given by the more recent
proposal of deformed (or doubly) special relativity (DSR) [4]. This theory is based
on the generalization of the standard energy-momentum dispersion law of particles
$P^2=m^2$. The deformation is achieved by modifying the action of the Lorentz group
on momentum space by means of a new observer-independent constant
$\k$, with the dimensions of energy
(usually identified with the Planck energy).
In this framework, the \tls of momenta become nonlinear, and that of positions
momentum dependent. Special relativity is recovered in the limit $\k\to0$.
Different choices of the deformed dispersion law correspond to different DSR models
and, even imposing suitable physical constraints, there exist in principle infinite
inequivalent models.
The physical motivations for the introduction of DSR are given by the possibility of
explaining some anomalies observed in high-energy cosmic ray distribution [5]
by means of deformed dispersion relations and by the theoretical requirement that
the Planck energy, which sets the scale for quantum gravity,
be invariant under Lorentz transformations\footnote{$^\ddagger$}
{\small More precisely, in most DSR models, the energy $\k$ is not left invariant by the
deformed transformations, but sets an observer-independent limit on the energy or
momentum of particles.
An extreme example of this fact is given by the Snyder model [6], where Lorentz
\trans act in the canonical (linear) way, and only the action of the translation
generators is nonlinear.}.

Algebraically, DSR theories can be realized in two equivalent ways,
either identifying the generators of the translations with the phase space momenta
and then deforming the \poi group, as in the $\k$-\mi algebra approach [7],
or by maintaining the form of the \poi algebra, but making it act nonlinearly on
momentum space, as in the case of the MS model [8]. The second approach is especially
convenient in the case of \des algebra, where, as mentioned above, even classically the
generators of the translations do not coincide with the canonical momenta.

From a physical perspective, the geometry of the spacetime on which DSR models
operate has a fundamental importance. Unfortunately, however, DSR models are
usually defined only in momentum space and the spacetime geometry is not fixed
uniquely from their postulates. Although it is possible to define DSR theories in
ordinary spacetime, their most natural realization appears nevertheless to be in
terms of noncommutative geometry, with momentum-dependent metric [9-11].
The momentum dependence of the metric has lead to a proposal for a generalization
of \GR that allows for the dependence of the geometry on the energy at which it is
probed [12].

The simplest way to construct a DSR model starting from canonical special relativity
was suggested in [13]: one can define the physical momenta as functions of auxiliary
variables which transform in the standard way under Lorentz transformations.
The deformed \tls and dispersion relations of the physical momenta then follow
from this definition.
More recently, it has been shown that also the definition of a suitable
non-commutative position space can be obtained by an analogous procedure [14].

Algebraically, \des space and the DSR momentum space have a very similar structure,
both being realized by imposing a quadratic constraint on the \coo of a
five-dimensional space [15].
However, their physical interpretation is different: \des space  has a
natural riemannnian structure, and one can choose arbitrary \coo on it;
momentum space has no such structure, and different parametrizations cannot be
interpreted as physically equivalent, unless further structure is added. In fact,
they lead to inequivalent DSR models with different dispersion relations.
Of course, a rigorous discussion of this topic requires a precise operational
definition of momentum measurements.

In this paper, we extend DSR models to the case of \des spacetime.
The first example of a DSR deformation of the \des algebra, limited to the momentum
sector of phase space, was given in [16]. Later, the authors of [17] gave a
different realization, extended to the full phase space. However, their approach
was purely algebraic, since they did not define a metric structure on \des space.
This may lead to ambiguities in the interpretation of the spacetime structure.

It is interesting to remark that the deformed \des algebra has two invariant scales,
beyond the speed of light. These are the \cosc $\L$ and the Planck energy $\k$
(or equivalently, the radius of curvature $\a\sim10^{25}$ m and the Planck length
$1/\k\sim10^{-35}$ m).
The two scales differ by 60 orders of magnitude and are related to the opposite
extrema of the range of observable physical phenomena. The origin of such
difference is not explained by modern physics. One of the models discussed in
this paper possesses a duality for the interchange of $\a$ and $\k$ together with
the interchange of positions and momenta.

The paper is organized as follows: in section 2 we discuss the \des algebra and
different parametrizations of \des space, realized as a hyperboloid embedded in
five-dimensional spacetime. In section 3 we discuss the dynamics of a free particle
in \des space. Section 4 and 5 are devoted to the study of the generalization of
the MS model to \des space. In section 6, a different generalization of DSR in \des
space is considered, related to the Snyder model. An alternative realization is
given in section 7. In section 8, some physical implications of our results are
discussed.

Although we shall not consider this issue in detail, all our result can be
straightforwardly extended to the \ads case, by simply changing the sign of the
cosmological constant.

\bigbreak
We use the following notations: $A,B=0,\dots,4$; $\m,\n=0,\dots,3$; $i,j=1,\dots,3$.
Except when dealing explicitly with the spacetime metric $g_\mn$, we always use
lower indices, for example $X_\m\id\y_\mn X^\m$,
where $X^\m$ are the natural (contravariant) coordinates.
The manipulation of indices are always performed
with the flat metric $\y_\mn={\rm diag}\ (1,-1,-1,-1)$, and not with the
metric $g_\mn$. The product between two 4-vectors
$\y_\mn V^\m W^\n=\y^\mn V_\m W_\n$
is denoted by $V\pun W$, and if $V=W$ by $V^2$.
For 5-vectors, we write the indices explicitly.
We also use coordinates without superscripts when dealing with expressions that
do not depend on the specific choice of coordinates.

\section{2. de Sitter space}
We review some properties of de Sitter space and its symmetry group,
which are not easily found in the literature. In particular, we discuss some
coordinate systems that will be useful in the following.
Unfortunately, contrary to Minkowski space, \des space does not admit a natural
choice of coordinates. In particular, as we shall see, different quantities
have simpler expressions in different coordinate systems. We shall therefore
alternate between them, depending on the subject under consideration.

\subsect{2.1. Generalities}
It is well known that \des space can be realized as a hyperboloid of equation
$\x_A^2=-\a^2$ embedded in 5-dimensional flat space, with coordinates $\x_A$ and
metric tensor $\y_{AB}={\rm diag}\ (1,-1,-1,-1,-1)$. In the following, we shall
often use the traditional notation $\L=1/\a^2$ for the cosmological constant.

The isometries of \des space are generated by the \des algebra. This can be
identified with the Lorentz algebra $so(1,4)$ of the 5-dimensional space, which
leaves invariant the hyperboloid. The generators $J_{AB}$ of the Lorentz algebra
read, in terms of the 5-dimensional canonical positions $\x_A$ and momenta $\p_A$,
$J_{AB}=\x_A\p_B-\x_B\p_A$, and obey the \pb
$$\{J_{AB},J_{CD}\}=\y_{BC}J_{AD}-\y_{BD}J_{AC}+\y_{AD}J_{BC}-\y_{AC}J_{BD}.
\eqno(2.1)$$
Their interpretation as generators of the \des algebra
is obtained by splitting them into Lorentz generators $J_\mn$ and translation
generators $T_\m=\sqrt\L\,J_{4\m}$. The de Sitter algebra can then be written as
$$\eqalignno{
&\{J_\mn,J_{\r\s}\}=\y_{\n\s}J_{\m\r}-\y_{\n\r}J_{\m\s}+\y_{\m\r}J_{\n\s}-
\y_{\m\s}J_{\n\r},&\cr
&\{J_\mn,T_\l\}=\y_{\m\l}T_\n-\y_{\n\l}T_\m,\qquad\{T_\m,T_\n\}=-\L J_\mn.&(2.2)}$$

The Lorentz subalgebra of the 4-dimensional \des algebra is identical to the
flat space Lorentz algebra, and therefore its generators can be realized in the
standard way in terms of the 4-dimensional coordinates $X_\m$ and their canonically
conjugate momenta $P_\m$, as $J_\mn=X_\m P_\n-X_\n P_\m$.
Thus the positions and the momenta obey the usual \tls under Lorentz
transformations,
$$\{J_{\mn},X_\l\}=\y_{\m\l}X_\n-\y_{\n\l}X_\m,\qquad
\{J_{\mn},P_\l\}=\y_{\m\l}P_\n-\y_{\n\l}P_\m.\eqno(2.3)$$

As we shall see, the realization of the translation generators $T_\m$ depends
instead on the specific choice of \coo on the hyperboloid.

\subsect{2.2. Natural \coo}
The \des hyperboloid can be parametrized by arbitrary coordinates and, contrary
to the case of flat space, there is no privileged system of coordinates for \des
space. The systems commonly used in the applications to general relativity single
out the time coordinate, while the most interesting for our purposes are isotropic
in space and time. Since these systems of \coo are not very well known,
we shortly review their properties.

We start by considering the natural parametrization, given by $\hX_\m=\x_\m$.
The metric induced on the hyperboloid by the five-dimensional flat metric reads
$$\hat g_\mn=\y_\mn-{\L\,\hX_\m\hX_\n\over1+\L\hX^2},\qquad
\hat g^\mn=\y^\mn+\L\,\hX^\m\hX^\n.\eqno(2.4)$$
In these \coo no cosmological horizon arises at finite distance.

From the definition of $T_\m=\sqrt\L\,J_{4\m}$, it is easy to see that under
translations
$$\{T_\m,\hX_\n\}=-\sqrt{1+\L\hX^2}\ \y_\mn.\eqno(2.5)$$
The nontrivial effect of translations is of course due to the curvature of the space.

From (2.5) it is evident that the translation generators $T_\m$ do not coincide
with the momenta $\hP_\m=\p_\m$ canonically conjugate to $\hX_\m$.
In fact,
$$T_\m=\sqrt{1+\L\hX^2}\ \hP_\m,\eqno(2.6)$$
and the momenta transform as
$$\{T_\m,\hP_\n\}={\L\hX_\n\hP_\m\over\sqrt{1+\L\hX^2}}.\eqno(2.7)$$

\subsect{2.3. Conformal \coo}
The parametrization that yields the simplest form of the metric is given by
conformal coordinates $\tX_\m=2\x_\m/(1+\sqrt\L\,\x_4)$, with inverse $\x_\m=
\tX_\m/(1-\L\tX^2/4)$. In these \coo the metric takes the diagonal form
$$\til g_\mn ={\y_\mn\over(1-\L\,\tX^2/4)^2},\qquad\til g^\mn =\Con^2\y^\mn,
\eqno(2.8)$$
and displays a cosmological horizon at $\tX^2=4/\L$.

Under translations the position \coo transform as
$$\{T_\m,\tX_\n\}=-\left(1+{\L\tX^2\over4}\right)\y_\mn+{\L\over2}\,\tX_\m\tX_\n,
\eqno(2.9)$$
and hence, in terms of the canonical momenta $\tP_\m=(1+\sqrt\L\,\x_4)\,\p_\m/2$,
$$T_\m=\left(1+{\L\tX^2\over4}\right)\tP_\m-{\L\over2}\,\tX\pun\tP\,\tX_\m.
\eqno(2.10)$$
The momenta transform as
$$\{T_\m,\tP_\n\}=-{\L\over2}(\tX\pun\tP\,\y_\mn+\tX_\m\tP_\n-\tX_\n\tP_\m).
\eqno(2.11)$$

\subsect{2.4. Beltrami \coo}
Another useful parametrization of the \des hyperboloid is given by Beltrami
coordinates [2,3], $\bX_\m=\x_\m/\sqrt\L\,\x_4$, with inverse $\x_\m=\bX_\m/
\sqrt{1-\L\bX^2}$.
In these coordinates, the metric has the form
$$\bar g_\mn ={(1-\L\bX^2)\y_\mn+\L\bX_\m\bX_\n\over(1-\L\bX^2)^2},
\qquad\bar g^\mn=(1-\L\bX^2)\big(\y^\mn-\L\bX^\m\bX^\n\big).\eqno(2.12)$$
A cosmological horizon is present at $\bX^2=1/\L$.

Under translations, the \coo $\bX_\m$ transform as
$$\{T_\m,\bX_\n\}=-\y_\mn+\L\bX_\m\bX_\n.\eqno(2.13)$$
In terms of the canonical momenta $\bP_\m=\x_4\p_\m/\sqrt\L$, the translation
generators read
$$T_\m=\bP_\m-\L\,\bX\pun\bP\,\bX_\m,\eqno(2.14)$$
and
$$\{T_\m,\bP_\n\}=-\L(\bX\pun\bP\,\y_\mn+\bX_\m\bP_\n).\eqno(2.15)$$

\section{3. Motion in de Sitter space}
We consider now the motion of a free particle in de Sitter space in the coordinate
systems introduced in the previous section, using the hamiltonian formalism.
Equivalent results could be obtained with greater effort using the Dirac theory of
constrained systems.

\subsect{3.1. Generalities\footnote{\rm *}
{\small In this subsection we restore the difference between upper and lower indices.}}

The lagrangian of a free particle of mass $m$, invariant under \des transformations,
is given by
$$L={m\over 2}\ g_\mn \dX^\m\dX^\n,\eqno(3.1)$$
where a dot denotes derivative \wrt an evolution parameter $\t$.

Varying with respect to $X_\m$ one obtains the geodesics equations.
Alternatively, defining the canonically conjugate momenta
$$P_\m\id{\de L\over\de\dX^\m}=m\,g_\mn\dX^\n,\eqno(3.2)$$
(with inverse $\dot X^\m={1\over m}\,g^\mn P_\n$), one can define the Hamiltonian as
$$H=P_\m\dot X^\m-L={1\over2m}g^\mn P_\m P_\n.\eqno(3.3)$$

The \eom in hamiltonian form are
$$\dX^\m=\{X^\m,H\}=g^\mn P_\n,\qquad\dP_\m=\{P_\m,H\}=
-{\de g^{\l\n}\over\de X^\m}\ P_\l P_\n,\eqno(3.4)$$
and can be derived by varying the action
$$\int d\t\,(\dX^\m P_\m-H).\eqno(3.5)$$

An equivalent way to obtain the hamiltonian is to identify it
with the quadratic Casimir invariant of the \des group, $J_{AB}J_{AB}=
J_\mn J_\mn-{2\over\L}\ T_\m T_\m$, written in terms of the four-dimensional
phase space variables. It is easy to see that in this way one recovers the
previous results.

\subsect{3.2. Conformal coordinates}
These \coo give the simplest relation between velocity and momentum.
The hamiltonian takes the form (from now on we put $m=1$),
$$\til H={1\over2}\left(1-{\L\tX^2\over4}\right)^2\tP^2,\eqno(3.6)$$
with \fe
$$
\dot{\tX}_\m=\left(1-{\L\tX^2\over4}\right)^2\tP_\m,\qquad
\dot{\tP}_\m={\L\over2}\left(1-{\L\tX^2\over4}\right)\tP^2\,\tX_\m.\eqno(3.7)$$
In these coordinates, the 3-velocity ${\bf v}_i$ is given by
$${\bf v}_i\id{\dot{\tX}_i\over\dot{\tX}_0}={\tP_i\over\tP_0},\eqno(3.8)$$
and hence the relation between 3-velocity and momenta is the same
as in flat space.

The \fe may be integrated by substituting the first equation (3.7) into the
second. However, it is more
convenient to use the conservation law associated with the translations,
which gives a first integral $T_\m=A_\m$, with $A_\m$ a constant vector.
From (2.10) and (3.7),
$$\left(1+{\L\tX^2\over4}\right)\tP_\m-{\L\over2}\,\tX\pun\tP\,\tX_\m=
{(1+\L\tX^2/4)\dot{\tX}_\m-\L\,\tX\pun\dot{\tX}\,\tX_\m/2\over(1-\L\tX^2/4)^2}
=A_\m.\eqno(3.9)$$
Inverting, one obtains
$$\dot{\tX}_\m={1-\L\tX^2/4\over1+\L\tX^2/4}\left[\left(1-{\L\tX^2\over4}\right)
A_\m-{\L\over2}\,A\pun\tX\,\tX_\m\right],\eqno(3.10)$$
and therefore
$${\bf v}_i={(1-\L\tX^2/4)A_i-\L\,A\pun\tX\,\tX_i/2
\over(1-\L\tX^2/4)A_0-\L\,A\pun\tX\,\tX_0/2}.\eqno(3.11)$$

\subsect{3.3. Beltrami coordinates}
These \coo have the nice property that 3-dimensional geodesics are straight
lines.
The hamiltonian takes the form
$$\bar H={1\over2}\ (1-\L\bX^2)[\bP^2-\L(\bX\pun\bP)^2],\eqno(3.12)$$
with \eom
$$\eqalignno{
&\dot{\bX}_\m=(1-\L\bX^2)(\bP_\m-\L\,\bX\pun\bP\,\bX_\m),&\cr
&\dot{\bP}_\m=\L\big[\big(\bP^2-\L(\bX\pun\bP)^2\big)\bX_\m+
(1-\L\bX^2)\bX\pun\bP\,\bP_\m)\big].&(3.13)}$$
Hence the 3-velocity can be written as
$${\bf v}_i\id{\dot{\bX}_i\over\dot{\bX}_0}={\bP_i-\L\bX\pun\bP\bX_i
\over\bP_0-\L\bX\pun\bP\bX_0}.\eqno(3.14)$$
Its form no longer coincides with its flat space analogous.

The equations (3.13) are rather involved, but one can exploit
the conservation law $\dot T_\m=0$ to obtain a first integral,
$$\bP_\m-\L\bX\pun\bP\bX_\m={\dot{\bX}_\m\over1-\L\bX^2}=A_\m,\eqno(3.15)$$
for constant $A_\m$.
Inverting, one obtains
$$\dot{\bX}_\m=(1-\L\bX^2)A_\m,\eqno(3.16)$$
and hence
$${\bf v}_i={A_i\over A_0}.\eqno(3.17)$$
Therefore, free particles have constant 3-velocity and their trajectories
in 3-space are straight lines.

\subsect{\noindent\it 3.4. Natural coordinates}
These \coo do not give rise to particularly simple expressions.
Therefore, we just summarize the main results.
The hamiltonian has the form
$$\hat H={1\over2}\ [\hP^2+\L (\hX\pun\hP)^2],\eqno(3.18)$$
and yields the \eom
$$\dot{\hX}_\m=\hP_\m+\L\,\hX\pun\hP\,\hX_\m,\qquad
\dot{\hP}_\m=-\L\,\hX\pun\hP\,\hP_\m,\eqno(3.19)$$
with 3-velocity
$${\bf v}_i={\hP_i+\L\hX\pun\hP\hX_i\over\hP_0+\L\hX\pun\hP\hX_0}.\eqno(3.20)$$

One can again exploit the conservation law $\dot T_\m=0$ to obtain a first
integral,
$$\hP_\m+\L\hX\pun\hP\hX_\m={(1+\L\hX^2)\dot{\hX}_\m-\L\hX\pun\dot{\hX}\hX_\m\over
\sqrt{1+\L\hX^2}}=A_\m.\eqno(3.21)$$
Inverting, one obtains $\dot{\hX}_\m$ and then
$${\bf v}_i={A_i-\L\,A\pun\hX\hX_i\over A_0-\L\,A\pun\hX\hX_0}.\eqno(3.22)$$

\section{4. The MS model in de Sitter space}
DSR theories in flat space can be implemented in two different ways.
One can either deform the \poi algebra [7,9], imposing nonlinear \pb
between the generators, or maintain the canonical form of the algebra, but modify
its action on the momentum variables [8,13].
The first approach has been considered in [16,17] in order to derive a deformed
\des algebra. However, for the discussion of the extension of DSR models to the
full phase space, especially in the case of a \des background, the second
approach appears to be more useful.

The MS model was introduced in [8] and is characterized by a deformed dispersion
relation $p^2/(1-p_0/\k)^2=m^2$. A remarkable property of this model is that the
Planck energy $\k$ is left invariant under the deformed Lorentz transformations.
The covariant realization of the model in a noncommutative position space was
discussed in [10,11].

In [14] it was observed that the representation of the MS algebra in phase space
can be obtained in a straightforward way from the \poi algebra acting canonically
on a space of \coo $X_\m$, $P_\m$,
by performing the substitution
$$X_\m=(1-p_0/\k)\,x_\m,\qquad P_\m={p_\m\over1-p_0/\k},\eqno(4.1)$$
with inverse
$$x_\m=(1+P_0/\k)\,X_\m,\qquad p_\m={P_\m\over1+P_0/\k}.\eqno(4.2)$$
Here $x_\m$, $p_\m$ are interpreted as physical observables, in contrast with
the auxiliary variables $X_\m$, $P_\m$.

The symplectic structure of phase space is then deformed and takes the form [10,11],
$$\eqalignno{
&\{x_0,x_i\}={x_i\over\k},\qquad\{x_i,x_j\}=0,\qquad\{p_0,p_i\}=\{p_i,p_j\}=0,&\cr
&\qquad\qquad\quad\{x_0,p_0\}=1-{p_0\over\k},\qquad\{x_i,p_j\}=-\d_{ij},&\cr
&\qquad\qquad\quad\{x_0,p_i\}=-{p_i\over\k},\qquad\quad\{x_i,p_0\}=0.&(4.3)}
$$
In particular, the coordinates $x_\m$ do not commute.

One can apply the same procedure in the \des case.
In this context it is useful to rewrite the \des algebra in the form
$$\eqalignno{
&\{N_i,N_j\}=\eps M_k,\quad\{M_i,N_j\}=\eps N_k,\quad\{M_i,M_j\}=\eps M_k,&\cr
&\qquad\qquad\quad\{T_i,T_j\}=-\L\,\eps M_k,\quad\{T_0,T_j\}=-\L\,N_j,&\cr
&\qquad\qquad\quad\{M_i,T_j\}=\eps T_k,\quad\{M_i,T_0\}=0,&\cr
&\qquad\qquad\quad\{N_i,T_j\}=\d_{ij}T_0,\quad\{N_i,T_0\}=T_i,&(4.4)}
$$
where $M_k=\ha\eps J_{ij}$ are the generators of rotations and $N_i=J_{0i}$ the
generators of boosts.

The \pb between phase space variables maintain the form (4.3).
Also the deformed action of the Lorentz subalgebra on coordinates and momenta
is the same as in the flat space MS model [11],
$$\eqalignno{
&\{M_i,x_j\}=\eps x_k,\qquad\qquad\quad\ \{M_i,x_0\}=0,&\cr
&\{N_i,x_j\}=\d_{ij}x_0+p_ix_j/\k,\qquad\{N_i,x_0\}=x_i+p_ix_0/\k.&\cr
&\{M_i,p_j\}=\eps p_k,\qquad\qquad\quad\ \ \{M_i,p_0\}=0,&\cr
&\{N_i,p_j\}=\d_{ij}p_0-p_ip_j/\k,\qquad\ \{N_i,p_0\}=p_i-p_ip_0/\k.&(4.5)}
$$
The action of translations on coordinates and momenta depends instead on the
specific coordinates chosen for \des space. For example, in the natural
parametrization,
$$\eqalignno{
&\{T_\m,\hx_\n\}=-\sqrt\hfact\left[\y_\mn-{\L\over\k}\ {\hx_0\hx_\n\hp_\m\over\hfact}
\right],&\cr
&\{T_\m,\hp_\n\}={\L\,(\hx_\n-\hp_\n\hx_0/\k)\,\hp_\m\over\sqrt\hfact}.&(4.6)}
$$

An interesting physical implication of this model is that the cosmological constant
becomes effectively energy dependent. Consider for example natural \coo and define,
in analogy with (4.2), $\hx_4=(1+\hP_0/\k)\,\hX_4$, with $\hX_4=\x_4$. Then
$\hx_A^2=-\a^2/(1-\hp_0/\k)^2\id-1/\L(\hp_0)$. In particular, for $\hp_0\to\k$,
$\L(\hp_0)\to0$, i.e.\ particles with energy close to the Planck energy do not
experience the curvature of spacetime.

\section{5. Dynamics of the MS model in de Sitter space}
Also the hamiltonian of a free particle can be obtained by
substituting (4.1) into the undeformed hamiltonian [14].
The \eom can then be obtained by taking into account the deformed symplectic
structure (4.4), namely,
$$\eqalignno
{&\dot x_0=\{x_0,H\}=\left(1-{p_0\over\k}\right){\de H\over\de p_0}-{p_i\over\k}\
{\de H\over\de p_i}+{x_i\over\k}\ {\de H\over\de x_i},&\cr
&\dot x_i=\{x_i,H\}=-{\de H\over\de p_i}-{x_i\over\k}\ {\de H\over\de x_0},&(5.1)}$$
and
$$\eqalignno{
&\dot p_0=\{p_0,H\}=-\left(1-{p_0\over\k}\right){\de H\over\de x_0},\cr
&\dot p_i=\{p_i,H\}={\de H\over\de x_i}+{p_i\over\k}\ {\de H\over\de x_0}.&(5.2)}$$
Equivalently, the Hamilton equations can be obtained by varying the action in which
the substitution (4.1) has been done.

For example, in conformal \coo the hamiltonian is given by
$$\til H=\til\D^2\tp^2,\eqno(5.3)$$
where
$$\til\D={1\over1-\tp_0/\k}-{\L\over4}\,(1-\tp_0/\k)\,\tx^2.
\eqno(5.4)$$
The Hamilton equations then read
$$\dot{\tx}_\m=\til\D^2\tp_\m+{\L\over2\k}\,(1-\tp_0/\k)\,\til\D\,\tp^2\tx_0\tx_\m,
\eqno(5.5)$$
and
$$\dot{\tp}_\m={\L\over2}\,(1-\tp_0/\k)\,\til\D\,\tp^2(\tx_\m-\tx_0\,\tp_\m/\k).
\eqno(5.6)$$
They can also be recovered from the action
$$\eqalignno{
I=&\int d\t\left[\dot{\tX}_\m\tP_\m-{1\over2}(1-\L\tX^2/4)^2\ \tP^2\right]&\cr
=&\int d\t\left[{\tp_\m\over1-\tp_0/\k}\ {d\over d\t}\big[(1-\tp_0/\k)\,\tx_\m\big]-
{1\over2}\til\D^2\tp^2\right].&(5.7)}$$

The Hamilton equations (5.5) have acquired complicated terms proportional to
$\L/\k$, and are no longer linear in the momentum, so that it is not easy to invert
them in order to obtain $\tp_\m$ in terms of $\dot{\tx}_\m$.
Because of this, it is difficult to obtain the \eom in second order form,
even using the conservation law for $T_\m$.

Moreover, the property that the velocity has the same expression as in the
undeformed case, valid for the MS model, does not extend to the \des case.
In fact, this property was proven in [18] to hold for position-independent
hamiltonians. If one wishes to maintain its validity, one should look for a different
deformation of the symplectic structure. For the same reason, contrary to flat space,
the evolution parameter $d\t$ cannot be identified with the line element invariant under
the deformed transformations, which reads
$$ds^2={d\tx^2\over\til\D^2}={(1-\tp_0/\k)^2\,d\tx^2\over\left[1-{\L\over4}\,
(1-\tp_0/\k)^2\,\tx^2\right]^2}.\eqno(5.8)$$
It is interesting to notice that the metric (5.8) exhibits a momentum-dependent
cosmological horizon at $\L\tx^2=4(1-p_0/\k)^{-2}$.
The dependence of the horizon on the momentum
is of course related to the momentum dependence of the \cosc discussed at the end of
previous section, and (5.8) can be considered an example of rainbow metric [12].

\bigskip
An analogous calculation can be performed in natural coordinates.
The deformed hamiltonian  is
$$\hat H={1\over2}\left[{\hp^2\over(1-\hp_0/\k)^2}+\L(\hx\pun\hp)^2\right],\eqno(5.9)$$
with Hamilton equations
$$\eqalignno{
&\dot{\hx}_\m=(1-\hp_0/\k)^{-2}\hp_\m+\L(1-\hp_0/\k)\,\hx\pun\hp\ \hx_\m,&\cr
&\dot{\hp}_\m=-\L(1-\hp_0/\k)\,\hx\pun\hp\ \hp_\m.&(5.10)}$$
Also in this case one finds the same problems as with conformal coordinates.
The same problems hold for Beltrami coordinates as well, in which the \eom are even more
involved.
In particular, it does not seem that the three-dimensional geodesics are still
straight lines in the deformed theory.

Finally, we notice that in the limits $\k\to\inf$ and $\L\to0$ one recovers the ordinary
\des space and the flat space MS model, respectively, while the limit $p_0\to\k$ is
analogous to that of the MS model [8].

\section{6. DSR in de Sitter space in a Snyder-like basis}

It is known that DSR theories can be realized in several different ways.
An interesting  realization is given by the so-called Snyder basis [6], which
is characterized by the dispersion relation $\cP^2/(1-\cP^2/\k^2)=m^2$, that
implies that the rest mass of particles must always be less than $\k$.
Another important property of this basis is that only the action of the translations is
deformed, while that of the Lorentz group is not affected. This example illustrates
the fact that the most relevant characteristic for the implementation of DSR is
the deformation of the action of translations (and hence a modified composition law
of momenta) and not that of Lorentz transformations, as usually postulated.

\subsect{6.1 Minkowski space}
Let us briefly review the case of flat spacetime. It is easy to see that,
in analogy with our previous treatment of the MS model, the easiest way to obtain the
Snyder realization of DSR is to define new coordinates from the canonical $X_\m$,
$P_\m$, which are thus interpreted as auxiliary variables,
$${\cal X}_\m=\snyd\,X_\m,\qquad{\cal P}_\m={P_\m\over\snyd},\eqno(6.1)$$
where $\O=1/\k^2$ is the Planck area\footnote{$^\sharp$}{\small In principle one may
choose  a negative sign for $\O$, obtaining an inequivalent model with rather
different properties [19].}. The inverse
\trans are
$$X_\m=\sqrt{1-\O\,\cP^2}\,\cX_\m,\qquad P_\m={\cP_\m\over\sqrt{1-\O\,\cP^2}},
\eqno(6.2)$$
One has then,
$$\{\cX_\m,\cX_\n\}=-\O(\cX_\m\cP_\n-\cX_\n\cP_\m),\qquad\{\cP_\m,\cP_\n\}=0,
\qquad\{\cX_\m,\cP_\n\}=\y_\mn-\O\,\cP_\m\cP_\n.\eqno(6.3)$$
The Lorentz transformations acting on $\cX_\m$, $\cP_\m$ maintain the canonical
form (2.3).
The translation generators $T_\m$ must instead be identified with
$P_\m=\cP_\m/\sqrt{1-\O\,\cP^2}$. Their action changes accordingly,
$$\{T_\m,\cX_\n\}={\y_\mn\over\snyi},\qquad\{T_\m,\cP_\n\}=0.\eqno(6.4)$$

The invariant hamiltonian for a free particle can be written as
$$H={P^2\over2}=\ha\ {\cP^2\over1-\O\,\cP^2},\eqno(6.5)$$
with \eom
$$\dot{\cX}_\m={\cP_\m\over1-\O\,\cP^2},\qquad\dot{\cP}_\m=0.\eqno(6.6)$$
It follows that $\dot\cX_\m=A_\m$ is constant.
The 3-velocity is then given by
$${\bf v}_i={\cP_i\over\cP_0}={A_i\over A_0},\eqno(6.7)$$
and the 3-dimensional geodesics are straight lines. Moreover, it is easy to verify
that the invariant line element $ds^2=(1-\O\,\cP^2)d\cX^2$ can be identified with
$d\t^2$, with $\t$ the evolution parameter.

\subsect{6.2 \des space}
Let us now extend the above construction to the case of \des space in the Beltrami
\coo of section 2.4.
The substitution (6.1) yields
$$\eqalignno{
&\bcX_\m=\sqrt{1+\O\bP^2}\,\bX_\m=\sqrt{\O\,\p^2+(1+\L\x^2)^\mo}\ \x_\m,&\cr
&\bcP_\m={\bP_\m\over\sqrt{1+\O\bP^2}}={\p_\m\over\sqrt{\O\,\p^2+(1+\L\x^2)^\mo}},
&(6.8)}$$
where $\x_\m$ are as usual the coordinates of the five-dimensional embedding space.
Inverting,
$$\x_\m={\bcX_\m\over\F},\qquad\p_\m=\F\bcP_\m,\eqno(6.9)$$
where
$$\F=\sqrt{(\opb)^\mo-\L\bcX^2}.\eqno(6.10)$$
The phase space \coo $\bcX_\m$, $\bcP_\m$ satisfy the \pb (6.3).

In terms of the variables $\bcX_\m$, $\bcP_\m$, the Lorentz generators of the \des
algebra (2.2) have canonical form, while the translation generators read
$$T_\m={1\over\sqrt\opb}\ \big[\bcP_\m-\L(\opb)\bcX\pun\bcP\,\bcX_\m\big],
\eqno(6.11)$$
and
$$\eqalignno{
&\{T_\m,\bcX_\n\}=-{1\over\sqrt\opb}\ \big[\y_\mn-\L(\opb)^2\,\bcX_\m\bcX_\n+
\L\O(\opb)\bcX\pun\bcP\,\bcP_\m\bcX_\n)\big],&\cr
&\{T_\m,\bcP_\n\}=-\L\sqrt\opb\ \big[\bcX\pun\bcP(\y_\mn-\O\,\bcP_\m\bcP_\n)
-(\opb)\bcX_\m\bcP_\n\big].&(6.12)}$$

One can also define a hamiltonian, invariant under the full deformed \des group,
$$H=\ha\ \F^2\big[\bcP^2-\L(\opb)(\bcX\pun\bcP)^2\big].
\eqno(6.13)$$
Unfortunately, the Hamilton equations take an extremely involved form and we shall
not report them here. The invariant metric for this model is
$$g_\mn={\F^2\y_\mn+\L\bcX_\m\bcX_\n\over\F^4}.\eqno(6.14)$$

In the limits $\O\to0$ and $\L\to0$ one recovers the ordinary \des space and the
flat space Snyder model of previous section, respectively.
Also interesting is the presence of a cosmological horizon at $\L\bcX^2=1-\O\bcP^2$
in the metric (6.14), whose location is momentum dependent.
\bigbreak
\section{7. A different Snyder-like realization}
The Snyder realization of DSR in \des space given in the previous section is
rather awkward.
In this section, we consider a slightly different realization, which takes a
more symmetric form and gives rise to more elegant formulas.
The algebra of this model displays some similarities with that proposed in [17].

We define
$$\cbX_\m=\sqrt{1+\O\p^2\over1+\L\x^2}\ \x_\m\qquad
\cbP_\m=\sqrt{1+\L\x^2\over1+\O\p^2}\ \p_\m,\eqno(7.1)$$
with inverse
$$\x_\m=\sqrt{\op\over\lx}\ \cbX_\m,\qquad
\p_\m=\sqrt{\lx\over\op}\ \cbP_\m.\eqno(7.2)$$
The \coo (7.1) satisfy the \pb
$$\eqalignno{
&\{\cbX_\m,\cbX_\n\}=-{\O\,(\lx)\over\lox}\ (\cbX_\m\cbP_\n-\cbX_\n\cbP_\m),&\cr
&\{\cbP_\m,\cbP_\n\}=-{\L\,(\op)\over\lox}\ (\cbX_\m\cbP_\n-\cbX_\n\cbP_\m),&\cr
&\{\cbX_\m,\cbP_\n\}=\y_\mn-{\L\,(\op)\cbX_\m\cbX_\n+\O\,(\lx)\cbP_\m\cbP_\n\over
\lox}\,.&(7.3)}$$

The Lorentz generators of the \des algebra have canonical form, while the dilatation
generators are
$$T_\m=\sqrt{\lox\over\op}\ \cbP_\m,\eqno(7.4)$$
and their action is given by
$$\eqalignno{
&\{T_\m,\cbX_\n\}=-\sqrt{\lox\over\op}\left[\y_\mn-{\L(\op)\over\lox}
\left(\cbX_\m\cbX_\n+\O\,{(\lx)\cbX\pun\cbP\,\cbP_\m\cbX_\n\over\lox}\right)\right],
&\cr
&\{T_\m,\cbP_\n\}=-\L\sqrt{\op\over\lox}\
\left[\cbX_\m\cbP_\n-\cbX_\n\cbP_\m+\O\,{(\lx)\cbX\pun\cbP\,\cbP_\m\cbP_\n\over\lox}
\right].&(7.5)}$$

The hamiltonian of a free particle can be obtained from the Casimir invariant of the
\des algebra, and takes the form
$$H=\ha\left[{1-\L\cbX^2\over\op}\ \cbP^2+\L(\cbX\pun\cbP)^2\right].\eqno(7.6)$$
Taking into account the symplectic structure (7.3), the \eom ensuing from the
hamiltonian are
$$\eqalignno{
&\dot{\cbX}_\m=(\lx)\left[{\cbP_\m\over\op}-{\L\O\,\cbX\pun\cbP\,\cbP^2\cbX_\m
\over\lox}\right]\ ,&\cr
&\dot{\cbP}_\m={\L\O(\lx)\over\lox}\ \cbX\pun\cbP\,\cbP^2\cbP_\m.
&(7.7)}$$
Also in this case, there does not seem to exist a simple relation between velocity
and momentum.

The 4-dimensional metric can be derived in the usual way from the 5-dimensional flat
metric subject to the constraint
$$\x_4=\sqrt{1+\L\x^2}=\sqrt{\lox\over\lx},\eqno(7.8)$$
and reads
$$g_\mn={\op\over\lx}\left[\y_\mn+\L{1+\O(\lx)\cbP^2\over(\lx)(\lox)}\,\cbX_\m\cbX_\n
\right].\eqno(7.9)$$
Also in this case there is no evident relation between the metric and the differential
$d\t$ of the evolution parameter.
It is interesting to notice that, in addition to the cosmological horizon at
$\cbX^2=1/\L$, the metric (7.9) presents a second momentum-dependent coordinate
singularity at $\L\cbX^2=1/\O\cbP^2$, or better $\cbX^2\cbP^2=1/\L\O$.
However, for such values of $\cbX$ and $\cbP$ the model is ill-defined (see (7.3)):
this region is also far beyond the range of physically observable phenomena, since
$1/\L\O\sim10^{120}$.

In the limit $\L\to0$ one of course recovers the flat-space Snyder model of previous
section, while in the limit $\O\to0$ one gets the standard \des space, although with
noncanonical \pb between positions and momenta (since the momenta are identified with
the translation generators in this limit).
More interesting are the limits $\cbX\to\a$ and $\cbP\to\k$. For $\cbX\to\a$, one is
close to the cosmological horizon, and the symplectic structure reduces to the
undeformed one obtained in the limit $\O=0$.
The limit $\cbP\to\k$ corresponds instead to the extremal value of the momentum.
In this limit, the symplectic structure is that of the flat Snyder model, $\L=0$,
and the metric and the hamiltonian are singular.

It is also interesting to notice that the \pb (7.3) lead after quantization to
generalized commutation relations of the most general kind proposed in [20] that,
in case of negative $\L$ and $\O$, imply the existence of both a minimal length and
momentum.

Another interesting property of this model is the existence of a duality for the
exchange of $\cbX\leftrightarrow\cbP$, together with $\L\leftrightarrow\O$.
This duality connects the high-energy/short-distance regime, governed by the Planck
area $\O$, with the low-energy/long-distance regime, governed by the cosmological
constant $\L$.
\vfill\eject
\section{8. Conclusions}
It is known that DSR models can be derived from a 5-dimensional momentum
space of coordinates $\p_A$, subject to the constraint $\p_A^2=-\k^2$ [15].
This is similar to the \des constraint for the spacetime coordinates.
However, the physical interpretation is quite different.
First of all, \des spacetime inherits a metric structure from the
5-dimensional space and this allows one to define a curvature. Different
systems of \coo are physically equivalent.
The momentum space, instead, does not possess a metric structure and its
\coo cannot be considered physically equivalent, unless one adds further
structure. In fact, different realizations of DSR lead to different physical
theories. Moreover, the mere existence of a \des group of transformations
on a four-dimensional manifold does not automatically imply that this can be
identified with \des space.

With these remarks in mind, one may try to construct a realization of a deformed
\des relativity starting from five-dimensional space, similarly to what has been
done for flat space [21]. Unfortunately, however,
it is not possible to impose contemporary constraints on the five-dimensional
positions and momenta, and one is forced to start from a six-dimensional space.
The construction of a hamiltonian formalism in six-dimensional
phase space with \coo $\X_M$ and momenta $\P_M$, subject
to the constraints $\X_M^2=-\a^2$, $\P_M^2=-\k^2$ will be the subject of a separate
paper [19].

From the study of DSR in \des space one can also learn some lessons concerning the
flat space limit. First of all, it is useful to distinguish the translation
generators, that dictate the conservation laws for the momentum, from the physical
momentum, identified with the phase space momentum variables. This observation
also gives a
physical meaning to the auxiliary variables obeying canonical \tls introduced in
ref.\ [13], whose interpretation was unclear: they are simply the generators of
translations. Moreover, it appears that the distinguishing feature of DSR is not
the deformation of the Lorentz symmetry, as usually postulated, but rather that of
the translation symmetry, as shown by the Snyder model discussed in section 6 and 7.
Of course, a complete discussion of this topic requires an operational definition
of the momentum of a particle.

Although we have not considered this subject in detail,
it is also important to stress that all our considerations can be easily extended
to the case of \ads space, by simply changing the sign of the cosmological constant
$\L$.

\vfill\eject
\beginref
\ref [1] M. Tegmark et al, \PR{D69}, 103501 (2004).
\ref [2] R.L. Mallett and G.N. Fleming, \JMP{14}, 45 (1973).
\ref [3] H.Y. Guo, C.G. Huang, Z. Xu and B. Zhou, \PL{A331}, 1 (2004).
\ref [4] G. Amelino-Camelia, \PL{B510}, 255 (2001), \IJMP{D11}, 35 (2002).
\ref [5] G. Amelino-Camelia and T. Piran, \PR{D64}, 036005 (2001).
\ref [6] H.S. Snyder, \PR{71}, 38 (1947);
J. Kowalski-Glikman, S. Nowak, \IJMP{D13}, 299 (2003).
\ref [7] J. Lukierski, H. Ruegg and W.J. Zakrzewski, \AoP{243}, 90 (1995).
\ref [8] J. Magueijo and L. Smolin, \PRL{88}, 190403 (2002).
\ref [9] J. Kowalski-Glikman, \MPL{A17}, 1 (2002).
\ref [10] A. Granik, \hep{0207113}.
\ref [11] S. Mignemi, \PR{D68}, 065029 (2005).
\ref [12] J. Magueijo and L. Smolin, \CQG{21}, 1725 (2004).
\ref [13] S. Judes and  M. Visser, \PR{D68}, 045001 (2003).
\ref [14] S. Ghosh, P. Pal, \PR{D75}, 105021 (2007).
\ref [15] J. Kowalski-Glikman, \PL{B547}, 291 (2002);
J. Kowalski-Glikman, S. Nowak, \CQG{20}, 4799 (2003).
\ref [16] S. Mignemi, \MPL{A18}, 643 (2003).
\ref [17] J. Kowalski-Glikman and L. Smolin, \PR{D70}, 065020 (2004).
\ref [18] S. Mignemi, \IJMP{D15}, 925 (2006).
\ref [19] S. Mignemi, in preparation.
\ref [20] A. Kempf, G. Mangano and R.B. Mann, \PR{D52}, 1108 (1995).
\ref [21] F. Girelli, T. Konopka, J. Kowalski-Glikman, E.R. Livine,
\PR{D73}, 045009 (2006)
\endref
\end